\documentclass[twocolumn,showpacs,preprintnumbers,amsmath,
amssymb,aps,prl]{revtex4}

\usepackage{epsfig}
\usepackage{color}
\usepackage{latexsym}
\usepackage{bm}
\usepackage{times}

\makeatletter

\newcommand{\cN}{{\cal N}}

\newcommand{\dg}{\delta}

\newcommand{\eg}{\epsilon}

\newcommand{\lam}{\lambda}

\newcommand{\di}{\partial} 
\newcommand{\be}{\begin{equation}} 
\newcommand{\ee}{\end{equation}} 
\newcommand{\bearr}{\begin{eqnarray}}
\newcommand{\eearr}{\end{eqnarray}} 

\newcommand{\bftheta}{\bm{\theta}} 
 
\newcommand{\Cyc}[1]{
  \vtop{\mathsurround=0pt
  \ialign{##\crcr$\textstyle{\rm Cyc}\strut$\crcr
    \noalign{\kern-0.4ex\nointerlineskip}{\tiny#1}\crcr}}\ }

\begin{document} 

\title{The Poisson bracket on free null initial data for gravity}
\author{\bf Michael P. Reisenberger}
\affiliation{Instituto de F\a'{\i}sica, Facultad de Ciencias,
        Universidad de la Rep\a'ublica,
        Igu\a'a 4225, CP 11400 Montevideo, Uruguay}
\date{November 17, 2008}

\begin{abstract}
Free initial data for general relativity on a pair of intersecting null 
hypersurfaces are well known, but the lack of a Poisson bracket and concerns 
about caustics have stymied the development of a constraint free canonical 
theory. Here it is pointed out how caustics and generator crossings can be 
neatly avoided and a Poisson bracket on free data is given. On sufficiently 
regular functions of the solution spacetime geometry this bracket matches 
the Poisson bracket defined on such functions by the Hilbert action via 
Peierls' prescription. The symplectic 2-form is also given in terms of free 
data.
\end{abstract}



\maketitle

\begin{figure}
\begin{center}

\includegraphics{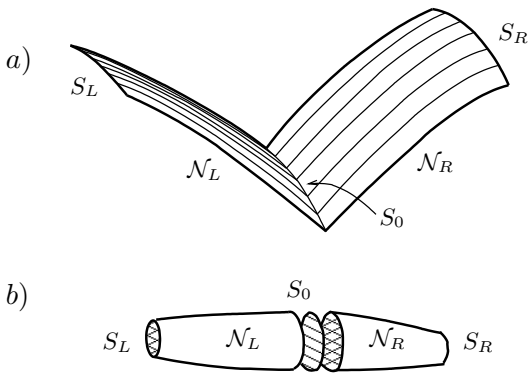}
\caption{a) A double null sheet in 2+1 dimensional spacetime. b) In 3+1 
dimensional spacetime $\cN$ is a 3-manifold consisting of two solid cylinders
joined on a disk, here shown without regard to their embedding in spacetime.}
\label{Nfigure}
\end{center}
\end{figure}

A constraint free canonical formulation of general relativity (GR) 
is of interest not least because at present the handling of constraints 
absorbs most of the effort invested in canonical approaches to quantizing 
gravity. Already in the 1960s free initial data for GR were identified on 
certain types of piecewise null hypersurfaces \cite{Sachs,Penrose,Bondi,
Dautcourt}, in particular on a ``double null sheet''. This is a compact 
hypersurface $\cN$ consisting of two null branches, $\cN_L$ and $\cN_R$, 
swept out by the two future directed normal congruences of null geodesics 
(called {\em generators}) emerging from a spacelike 2-disk $S_0$, the 
branches being truncated on disks $S_L$ and $S_R$ before the 
generators form caustics (see Fig.\ \ref{Nfigure}).



Nevertheless a constraint free canonical theory was not constructed, for two 
reasons: First, the Poisson brackets of the free initial data were unknown.
Second, in order that $\cN$ not enter its own future, implying intractable 
constraints on the otherwise free initial data, the generators must not cross 
at interior points of $\cN$ \cite{Wald}. But excluding such crossings itself seemed to require intractable conditions on the data.
Here a Poisson bracket on free data corresponding to the Hilbert action is 
presented, and a simple way to avoid caustics and generator crossings is pointed 
out.



The resulting framework seems ideal for attempting a semi-classical proof of the
Bousso entropy bound
\cite{Bekenstein,tHooft,Susskind,Bousso} in the vacuum gravity case, since a 
branch $\cN_A$ ($A = L\ \mbox{or}\ R$) of $\cN$ is a ``light sheet'' in the 
terminology of Bousso \cite{Bousso} provided the generators are not diverging 
at $S_0$. 




Canonical GR using {\em constrained data} on double null sheets
has been developed by several authors \cite{Torre, Goldberg1, Goldberg2, 
Vickers}. Presumably the present Poisson brackets can be interpreted as Dirac 
brackets in those frameworks. Results on the brackets of part of the free 
data are given in Refs. \cite{Goldberg2, GR}. Reference \cite{GR} gives
perturbation series in Newton's constant for the brackets of free data living 
on the bulk of $\cN$ consistent with the present work, but no brackets of the 
surface data on $S_0$. Reference \cite{Goldberg2} presents distinct free data 
on the bulk of $\cN$, which are claimed to form a canonically conjugate pair on 
the basis of a machine calculation of Dirac brackets. It would be interesting 
to see if they are conjugate according to the bracket obtained here.




A special chart $(v_A, \theta^1, \theta^2)$ will be used on each branch 
$\cN_A$ of $\cN$, with $v_A$ a parameter along the generators and $\theta^a$ 
($a = 1,2$) constant along these. Since $\di_{v_A}$ is tangent to the 
generators it is null and normal to $\cN_A$. The line element on $\cN_A$ 
thus takes the form
\be   \label{line_element}
ds^2 = h_{ab} d\theta^a d\theta^b,
\ee
with no $dv$ terms. $v_A$ is taken proportional to the square root of 
$\rho \equiv \sqrt{\det h}$, the area density in $\theta$ coordinates on 2D 
cross sections of $\cN_A$, and normalized to $1$ at $S_0$. Thus 
$\rho = \rho_0(\theta^1, \theta^2) v_A^2$, with $\rho_0$ the area density on 
$S_0$. Any affine parameter $\eta$ on the generators is related to $v$ by 
\cite{MR}
\be \label{focusing}
0 = R[\di_v,\di_v] = \frac{2}{v}\frac{d}{dv}\ln \Big|\frac{d\eta}{dv}\Big| 
+ \frac{1}{4}\di_v e_{ab}\di_v e^{ab},
\ee 
a vacuum Einstein equation equivalent to the ``focusing equation'' (9.2.32 in \cite{Wald}). Here 
$e_{ab} = h_{ab}/\rho$, a unit determinant, symmetric $2\times 2$ matrix.

 

 


At caustic points $v^2 \equiv \rho/\rho_0$ vanishes, so the caustic {\em free} 
$\cN$ are represented by initial data on coordinate domains in which $v >0$.
In the absence of caustics generators can still cross on $\cN$ but the crossing 
points can be ``unidentified'': A spacetime which is locally 
isometric to a neighborhood of $\cN$, but in which the generators do not cross,
may be constructed by pulling the metric of the original spacetime back to the
normal bundle of $S_0$ using the exponential map 
(Ref. \cite{MR}, Appendix B).
The exclusion of caustics and crossings thus requiers no restriction on the data
at $v >0$. In particular it does not restrict the scope of the present work to 
weak fields.






Since $\cN_A$ is caustic free $d\theta^1\wedge d\theta^2$
is degenerate only along generators. $(v_A, \theta^1, \theta^2)$
is thus a good chart provided $dv_A \neq 0$ on the generators.
For smooth $S_0$, in a smooth vacuum solution, this is so if the
generators are converging everywhere on $S_0$ ($v$ decreasing away from $S_0$),
since by the focusing equation (\ref{focusing}) $v$ continues to decrease 
until a caustic is reached, and also if the generators are diverging everywhere
on $S_0$ but are truncated before they begin to reconverge. In these cases 
$e_{ab}$ induced by the spacetime geometry is smooth in $v$ and $\theta^a$.
Conversely, if (A) $v$ is not strictly constant on a generator, and (B) $e_{ab}$ is 
smooth in $v$, then by (\ref{focusing}) $dv \neq 0$ along the generator. The 
initial data will satisfy conditions A and B on all generators, ensuring that
$(v, \theta^1, \theta^2)$ is a good chart.

Sachs \cite{Sachs} showed (modulo convergence issues) that $e_{ab}$, specified 
on $\cN$ as a function of 
an affine parameter on the generators, 
together with additional data on $S_0$, is free initial data determining 
the geometry of a spacetime region to the future of $\cN$.
Here we assume that any Sachs data without caustics determines a unique 
maximal Cauchy development $D[\cN]$ of all of $\cN$, and that if the data depend 
smoothly on a parameter the solution does as well. Existence, uniqueness, and 
smooth dependence on parameters have been proved rigorously in a neighborhood of 
$S_0$ \cite{Rendall}.



We will use similar data, including $e_{ab}$, given on $\cN$ as a 
smooth function of $v$; and $\rho_0$, $\lam = - \ln|n_L\cdot n_R|$, and
\be
\tau_a = \frac{n_L\cdot\nabla_a n_R - n_R\cdot\nabla_a n_L}{n_L\cdot n_R}
\ee
specified on $S_0$. [Here $n_A = \di_{v_A}$ is the tangent to the generators of $\cN_A$, 
and inner products ($\cdot$) are evaluated using the spacetime metric.] $v_A$
ranges from $1$ on $S_0$ to $\bar{v}_A(\theta)$ on $S_A$. $\bar{v}_A(\theta)$, which
is another datum, is required to be $>0$ and $\neq 1$. The data are smooth 
functions of the $\theta^a$, which range over the unit disk 
$(\theta^1)^2 + (\theta^2)^2 \leq 1$.
Any valuation of these data determine, via (\ref{focusing}), Sachs data
free of caustics and thus, according to our asumptions, a solution to GR unique up to 
diffeomorphisms \cite{MR}.



However, because $\cN$ has a boundary, not all infinitesimal diffeomorphisms
are degeneracy vectors of the symplectic 2-form on $\cN$ \cite{MR}. Two further
data on $S_0$, $s^m_L$ and $s^i_R$, measure diffeomorphisms which are non-gauge
in this sense. $y_A^k = s_A^k(\theta)$ is the position of the endpoint on
$S_A \subset \di\cN$ of the generator $(\theta^1, \theta^2)$, in a fixed chart
$y_A$ on $S_A$. Since $s_A$ may be varied independently of the other data by
diffeomorphisms the complete data set is still free. We shall return to the
question of the significance of the $s_A$.



In sum, the data consist of 10 real $C^\infty$ functions, $\rho_0$, $\lam$, 
$\tau_a$, $\bar{v}_A$, and $s_A^k$, on the unit 2-disk with $\bar{v}_A > 0$ 
and $\neq 1$, and two $C^\infty$, real, symmetric, unimodular $2 \times 2$ 
matrix valued functions ($e_{ab}$ on $\cN_L$ and $\cN_R$) on the domains 
$\{(\theta^1)^2 + (\theta^2)^2 \leq 1, \min(1,\bar{v}_A(\theta)) \leq v_A 
\leq \max(1,\bar{v}_A(\theta))\}$, $A = L,R$ which match at $v_L = v_R = 1$ 
(i.e. on $S_0$). Our phase space is the space of valuations of these data.

An alternative representation of $e_{ab}$ can be obtained by expressing the 
degenerate line element (\ref{line_element}) on $\cN$ in terms of the complex 
coordinate $z = \theta^1 + i\theta^2$:
\bearr
       ds^2 & = & h_{ab}d\theta^a d\theta^b  \nonumber\\
	    & = & \rho (1 - \mu\bar{\mu})^{-1}[dz + \mu d\bar{z}][d\bar{z} + \bar{\mu}dz],
\label{mu_parametrization}
\eearr
with $\mu$ a complex number valued field of modulus less than 1 (sometimes 
called the {\em Beltrami differential}). $\mu$ encodes the two real degrees of 
freedom of $e_{ab} = h_{ab}/\rho$. This parametrization of $e_{ab}$ also works 
when $e_{ab}$ is not real, but then $\mu$ and $\bar{\mu}$ are no longer complex
conjugates.


Finding the Poisson bracket on initial data by inverting the (gauge fixed) 
symplectic 2-form, or other conventional approaches, turns out to be difficult.
But what is ultimately required of the bracket is that it gives the 
correct Poisson brackets for observables. We shall content ourselves with 
finding a bracket that satisfies this criterion. Observables will be defined 
as diffeomorphism invariant functionals of the metric $F[g]$ with $C^\infty$ 
functional derivatives $\dg F/\dg g_{\mu\nu}$ of compact support contained in 
the interior of the Cauchy development $D[\cN]$. The Poisson brackets of 
these observables may be defined via Peierls' \cite{Peierls} covariant formula 
in terms of the action and Green's functions \cite{MR}. 




To match the Peierls bracket on observables $\{\cdot,\cdot\}_\bullet$ need only
be {\em almost} inverse to the symplectic 2-form $\omega_\cN$ on $\cN$. Specifically, 
let $g$ be a metric satisfying the field equations and let $L^0_g$ be the space of 
perturbations of the metric that satisfy the field equations linearized about 
$g$ and vanish in a neighborhood of $\di\cN$.
Then $\{\cdot,\cdot\}_\bullet$ matches the Peierls bracket on observables (at $g$) if 
\be	\label{bracket_condition1}
        \dg \varphi = \omega_\cN[\{\varphi,\cdot\}_\bullet,\dg]\ \ \
\forall\:\dg \in L_g^0.
\ee
for all integrals, $\varphi$, of initial data against smooth test functions on
$\cN$ that vanish in a neighborhood of $\di\cN$ \cite{MR}.


We will use the symplectic 2-form of the Hilbert action, and we will
require $\{\cdot,\cdot\}_\bullet$ to be causal:
data at $p \in \cN$ must commute with data outside the causal domain of influence 
of $p$, domain which on $\cN$ reduces to just the generator(s) through $p$ \cite{MR}.
Then $\omega_\cN[\{\varphi,\cdot\}_\bullet,\dg]$ in (\ref{bracket_condition1}) can 
be expressed in terms of the initial data as $\Omega_\cN[\{\varphi,\cdot\}_\bullet,\dg]$ 
where $\Omega_\cN = \Omega_L + \Omega_R$ with \cite{MR}
\bearr	 
16\pi G \Omega_A[\dg_1,\dg_2] & = &\int_{S_0} d^2\theta 
\Big\{\dg_1\lam \dg_2\rho_0 + \dg_1\tilde{\tau}_{A\,k}\dg_2 s_A^k \nonumber\\
&& + \frac{1}{4}\dg_1^y \rho_0 \di_v e_{ab} \dg^y_2 e^{ab} \nonumber\\
&& + \frac{1}{2} \rho_0 \int_1^{\bar{v}} v^2 
\dg^\circ_1 e^{ab}\di_v \dg^\circ_2 e_{ab} dv\:\Big\} \nonumber\\ 
&& - (1 \leftrightarrow 2). \label{Omega_free_data}
\eearr
Here $\tilde{\tau}_{R\,i} dy_R^i = \rho_0[d \lam - \tau]$ and
$\tilde{\tau}_{L\,m} dy_L^m = \rho_0[d \lam + \tau]$;
$\dg^y = \dg - {\pounds}_{\xi_\perp}$, where $\xi_\perp 
= \dg s_A^k \di_{y_A^k}$ and the partial derivative $\di_{y_A^k}$ is taken at 
constant $v$; and $\dg^\circ = \dg^y - \frac{1}{2}\dg^y \ln \rho_0\, v \di_v$. 
In calculating (\ref{Omega_free_data}) no boundary terms were added to the 
Hilbert action because the Peierls bracket, which determines the brackets of 
observables, is unaffected by such terms. Equation (\ref{Omega_free_data}) is consistent
with \cite{Epp}.

Equation (\ref{bracket_condition1}) does not determine the bracket uniquely, nor
does it guarantee that it satisfies the Jacobi relations. Here a unique Poisson
bracket is obtained by defining a set $C$ of variations of the data containing 
those corresponding to spacetime metric variations in $L_g^0$ and imposing 
\bearr	
        \dg \varphi = \Omega_\cN[\{\varphi,\cdot\}_\bullet,\dg]\ \ \ 
\forall\:\dg \in C
	\ \ \ \mbox{and}\ \ \ \{\varphi,\cdot\}_\bullet \in C,\nonumber\\
&&
\label{final_bracket_condition}
\eearr
for all $\varphi$ obtained by smearing data with {\em any} $C^\infty$ test function 
on $\cN$ or $S_0$. The first condition 
ensures agreement with the Peierls bracket; the second that the Jacobi relations hold.


$C$ consists of all {\em complex} variations $\dg$ of the data such that 
(A) $\dg \bar{\mu}$ is smooth on $\cN$ while $\dg \mu$ is smooth on
$\cN_L - S_0$, $\cN_R - S_0$, and $S_0$, with possible jump dicontinuities
between them, and (B) $\dg$ leaves invariant on $S_A$ both $\rho_A$, the area density in 
the $y_A$ chart, and $\mu_A$, the Beltrami differential in the complex chart 
$y^1_A + i y^2_A$. ($\dg\bar{\mu}_A$ need {\em not} vanish on $S_A$.)


The use of the space $C$ of complex perturbations on the real phase space is 
strange but seems difficult to avoid.
Note however that all hamiltonian vectors defined by the 
$\bullet$ bracket satisfying (\ref{final_bracket_condition})
preserve the reality of observables, and of the metric on the interior of 
$D[\cN]$.


Because of the identity
\be	\label{omega_constraint}
\tilde{\tau}_{R\,i}\di_a s_R^i + \tilde{\tau}_{L\,m}\di_a s_L^m  
= 2\rho_0 \di_a \lam
\ee
$\Omega_\cN$ is degenerate with respect to variations of the data due 
to diffeomorphisms of the $\theta$ chart on $S_0$. This degeneracy can be 
removed by extending the phase space by making 
$\tilde{\tau}_R$ and $\tilde{\tau}_L$ independent, and treating 
(\ref{omega_constraint}) as a constraint (which generates diffeomorphisms of 
$\theta$ \cite{MR}). Then (\ref{final_bracket_condition}) defines the $\bullet$
brackets of the data uniquely as two point distributions on $S_0$ and $\cN$.


[A description without {\em any} constraint may be obtained by eliminating 
$\tilde{\tau}_L$ via (\ref{omega_constraint}), and $s_L$ via the gauge fixing 
$\theta = y_L$. The Dirac brackets of the remaining data are then identical to 
their extended phase space $\bullet$ brackets \cite{MR}.]


Solving (\ref{final_bracket_condition}) by a procedure like that of \cite{MR} 
yields:
\be
\{\rho_0(\bftheta_1),\lam(\bftheta_2)\}_\bullet 
 = 8\pi G \dg^2(\bftheta_2 - \bftheta_1),	
\label{lam_rho_bracket}
\ee
\be
\{s_A^i(\bftheta_1),\tilde{\tau}_{B\,j}(\bftheta_2)\}_\bullet 
 = 16\pi G \dg_{AB}\dg_j^i\dg^2(\bftheta_2 - \bftheta_1),	
\label{omega_s_R_bracket}
\ee
and $s_A$ and $\rho_0$ commute with all other data.

The brackets between $\tilde{\tau}_R$, $\tilde{\tau}_L$, and $\lam$ are 
\be
\{\lam(\bftheta_1),\lam(\bftheta_2)\}_\bullet = 0, \label{lam_lam_brack}
\ee
\bearr
\{\lam(\bftheta),\tau_R[f]\}_\bullet & = & 
8\pi G \Big[\frac{{\pounds}_f \mu}{(1 - \mu\bar{\mu})^2}
(\di_{v_R} \bar{\mu} - \di_{v_L} \bar{\mu}) \Big]_{\bftheta}, \qquad\nonumber\\
&&
\label{lam_omega_R_bracket}
\eearr
\bearr
\{\tau_R[f_1],\tau_R[f_2]\}_\bullet & = & - 16\pi G
\int_{S_0} \frac{1}{(1 - \mu\bar{\mu})^2}{\pounds}_{f_1} \mu 
(\eg {\pounds}_{f_2}\bar{\mu} \nonumber\\
&& - {\pounds}_{f_2}\eg\di_{v_R}\bar{\mu}) 
- (1 \leftrightarrow 2),
\label{omega_omega_R_bracket} 
\eearr
\bearr
\{\tau_R[f],\tau_L[g]\}_\bullet & = & 16\pi G \int_{S_0} 
\frac{1}{(1 - \mu\bar{\mu})^2}
{\pounds}_f \mu(\eg {\pounds}_g\bar{\mu} \nonumber\\
&& - {\pounds}_g\eg\di_{v_L}\bar{\mu}) 
- (f,R \leftrightarrow g,L), \label{omega_R_omega_L_bracket}
\eearr
%
the 
rest being obtainable from these by exchanging $L$ and $R$. 
$\eg$ is the area form $\rho_0 d\theta^1\wedge d\theta^2$, 
$\tau_R[f] = \int_{S_0} \tilde{\tau}_{R\,i} f^i d^2\theta$ with $f^i(\theta)$ 
test functions independent of the data, and $\tau_L[g]$ is defined similarly in
terms of test functions $g^m$. $f^i$ and $g^m$ define vector fields 
$f = f^i \di_{y_R^i}$ and $g = g^m \di_{y_L^m}$, and thus Lie derivatives.
The only unusual one is 
\be
{\pounds}_f \mu = f^z \di_z \mu + f^{\bar{z}} \di_{\bar{z}} \mu
- \mu [\di_z f^z - \di_{\bar{z}}f^{\bar{z}}] - \mu^2 \di_z f^{\bar{z}} 
+ \di_{\bar{z}} f^z
\nonumber
\ee


The brackets between $\mu$ and $\bar{\mu}$ are as follows: If ${\mathbf 1}, 
{\mathbf 2}$ 
denote the $(v, \theta)$ coordinates of any pair of points on $\cN$ 
\be   \label{mu_mu_brack}
0  =  \{\mu({\mathbf 1}),\mu({\mathbf 2})\}_\bullet 
= \{\bar{\mu}({\mathbf 1}),\bar{\mu}({\mathbf 2})\}_\bullet.
\ee
When ${\mathbf 1}$ and ${\mathbf 2}$ do not lie on the same branch also 
$\{\mu({\mathbf 1}), \bar{\mu}({\mathbf 2})\}_\bullet = 0$, but if ${\mathbf 1}$ and 
${\mathbf 2}$ do lie on the same branch, $\cN_A$, then
\bearr
\{\mu({\mathbf 1}),\bar{\mu}({\mathbf 2})\}_\bullet
& = & 4\pi G\, \frac{1}{\rho_0}\dg^2(\bftheta_2 - \bftheta_1)\,
H({\mathbf 1},{\mathbf 2})\Big[\frac{1 - \mu\bar{\mu}}{v_A}\Big]_{\mathbf 1}
\nonumber\\
&& \times\Big[\frac{1 - \mu\bar{\mu}}{v_A}\Big]_{\mathbf 2}\:
e^{\int_{\mathbf 1}^{\mathbf 2}
(\bar{\mu} d\mu - \mu d\bar{\mu})/(1 - \mu\bar{\mu})},\nonumber\\
&&
\label{mu_mubar_bracket}
\eearr
where if ${\mathbf 1}$ and ${\mathbf 2}$ lie on the same generator
the integral runs along the generator segment from ${\mathbf 1}$ 
to ${\mathbf 2}$, and $H$ is a step function equal to $1$ if ${\mathbf 1}$ 
lies on or between $S_0$ and ${\mathbf 2}$, and $0$ otherwise. 
Here and elsewhere the coefficient of $\dg^2(\bftheta_2 - \bftheta_1)$ 
is extended continuously to $\bftheta_2 \neq \bftheta_1$.



There remain the brackets of $\mu$ and $\bar{\mu}$ with the $S_0$ data $\lam$, 
$\tilde{\tau}_R$, and $\tilde{\tau}_L$. For ${\mathbf 1}$ on 
$\cN_R - S_0$ (i.e. $v_{R\,1} \neq 1$)
\be
\{\mu({\mathbf 1}),\lam({\bftheta_2})\}_\bullet 
= 4\pi G \frac{1}{\rho_0}\dg^2(\bftheta_2 - \bftheta_1)
[v_R \di_{v_R} \mu]_{\mathbf 1},
\label{mu_lam_bracket}
\ee
\be
\{\mu({\mathbf 1}), \tau_R[f]\}_\bullet = 8\pi G\Big[2{\pounds}_f \mu 
- \frac{{\pounds}_f \rho_0}{\rho_0} v_R\di_{v_R} \mu\Big]_{\mathbf 1},
\label{mu_omegaR_bracket}
\ee
\be
\{\mu({\mathbf 1}), \tau_L[g]\}_\bullet = 0,   \label{mu_omegaL_bracket}
\ee
while for ${\mathbf 1}$ on $S_0$
\be
\{\mu({\mathbf 1}),\lam({\mathbf 2})\}_\bullet = 0,
\label{mu_lam_0_bracket}
\ee
\be
\{\mu({\mathbf 1}), \tau_R[f]\}_\bullet 
= 8\pi G [{\pounds}_f \mu]_{\mathbf 1},  \label{mu_omega_R_0_bracket}
\ee
\be
\{\mu({\mathbf 1}), \tau_L[g]\}_\bullet 
= 8\pi G [{\pounds}_g \mu]_{\mathbf 1}.  \label{mu_omega_L_0_bracket}
\ee
On the other hand, for all ${\mathbf 1} \in \cN_R$ (including ${\mathbf 1} \in 
S_0$)
\bearr
\{\bar{\mu}({\mathbf 1}),\lam(\bftheta_2)\}_\bullet 
& = & 4\pi G \frac{1}{\rho_0}\dg^2(\bftheta_2 - \bftheta_1)\Big[(v_R \di_{v_R} 
\bar{\mu})_{\mathbf 1} \nonumber\\
&& + \Big(\frac{1}{v_R}\Big)_{\mathbf 1} e^{2\int_{\mathbf 1}^{\mathbf 2} 
(\mu d \bar{\mu})/(1 - \mu\bar{\mu})} (\di_{v_L}\bar{\mu})_{\mathbf 2}\Big],\nonumber\\
&&
\label{mubar_lam_bracket}
\eearr
\bearr
\{\bar{\mu}({\mathbf 1}),\tau_R[f]\}_\bullet & 
= & 8\pi G\Big[\Big(2{\pounds}_f \bar{\mu} - \frac{{\pounds}_f \rho_0}
{\rho_0} v_R\di_{v_R} \bar{\mu}\Big)_{\mathbf 1} \nonumber\\ 
&& - ({\pounds}_f \bar{\mu})_{{\mathbf 1}_0}
\Big(\frac{1}{v_R}\Big)_{\mathbf 1} e^{-2\int_{{\mathbf 1}_0}^{\mathbf 1}
(\mu d \bar{\mu})/(1 - \mu\bar{\mu})}\Big],\nonumber\\
&&
\label{mubar_omegaR_bracket}
\eearr
\bearr
\{\bar{\mu}({\mathbf 1}), \tau_L[g]\}_\bullet & = & 
8\pi G\Big[{\pounds}_g \bar{\mu} 
-\frac{{\pounds}_g \rho_0}{\rho_0}\di_{v_L} \bar{\mu}\Big]_{{\mathbf 1}_0}\nonumber\\ 
&& \times \Big(\frac{1}{v_R}\Big)_{\mathbf 1} 
e^{-2\int_{{\mathbf 1}_0}^{\mathbf 1} 
(\mu d \bar{\mu})/(1 - \mu\bar{\mu})},
\label{mubar_omegaL_bracket}  
\eearr
where ${\mathbf 1}_0 \in S_0$ is the base point of the generator through 
${\mathbf 1}$. Exchanging $L$ and $R$ in 
(\ref{mu_lam_bracket})--(\ref{mubar_omegaL_bracket}) 
gives the corresponding brackets for ${\mathbf 1}$ on $\cN_L$.

Finally, the brackets of $\bar{v}_A$ follow from the preceeding brackets and 
the fact that, by (\ref{final_bracket_condition}), $\rho_A = \rho_0 \bar{v}_A^2
|\det\di_a s_A^k|^{-1}$ at given $y_A$ commutes with everything. Alternatively,
$\bar{v}_A(\theta)$ may be replaced as a phase coordinate by $\rho_A(y_A)$.

Direct calculations confirm that these expressions for the brackets
satisfy the Jacobi relations, that they are invariant under diffeomorphisms
of the (arbitrarily chosen) coordinates $y_R^i$, $y_L^m$ and $\theta^a$ \cite{MR},
and that the constraint (\ref{omega_constraint}) generates diffeomorphisms of the 
$\theta^a$ \cite{MR}.


Strangely, the brackets do not preserve the reality of $e_{ab}$, i.e. the 
complex conjugacy of $\mu$ and $\bar{\mu}$.
An analytic functional $F$ of the data is real on real data iff it
equals $\bar{F}$, its {\em formal complex conjugate}, obtained by exchanging
$\mu$ and $\bar{\mu}$, leaving the $S_0$ data untouched, and replacing numerical
coefficients by their complex conjugates. $\{F,\cdot\}_\bullet$ preserves
the reality of the data for all formally real $F$ iff the bracket itself is real
in the sense that it equals the formal complex conjugate bracket
$\{\varphi,\chi\}_{\bullet c.c.} \equiv \overline{\{\bar{\varphi},
\bar{\chi}\}}_\bullet$.
But this is not so:
$\overline{\{\mu({\mathbf 1}),\bar{\mu}({\mathbf 2})\} _\bullet} 
\neq \{\bar{\mu}({\mathbf 1}),\mu({\mathbf 2})\}_\bullet$. Nevertheless, on
observables the bracket is real, as it reproduces the real Peierls bracket.
%
In fact, one may resolve the $\bullet$ bracket into (formal) real and imaginary 
parts, $\{\cdot,\cdot\}_\bullet = \{\cdot,\cdot\}_R + i \{\cdot,\cdot\}_I$, and
one finds that
\bearr
\{\cdot,\cdot\}_I &= &\frac{i}{8\pi G}\sum_{A = L, R} \int_{S_A} \eg
\frac{1}{(1 - \mu_A\bar{\mu}_A)^2}
[\{\cdot,\bar{\mu}_A\}_\bullet 
\overline{\{\bar{\mu}_A, \bar{\cdot}\}}_{\bullet} \nonumber\\ 
&& - \overline{\{\bar{\cdot},\bar{\mu}_A\}}_{\bullet}
\{\bar{\mu}_A,\cdot\}_\bullet].  \label{imaginary_part}
\eearr
In agreement with causality, $\{\bar{\mu}_A(q),\cdot\}_\bullet$ for 
$q \in S_A$ is a gravitational wave pulse that skims along $\cN_A$ without 
entering the interior of $D[\cN]$. It does not affect the metric there 
(Ref. \cite{MR}, Appendix C), so neither does $\{\phi,\cdot\}_I$ for any datum 
$\phi$.

$\{\cdot,\cdot\}_R$ 
is the pre-Poisson bracket 
$\{\cdot,\cdot\}_\circ$ of \cite{MR}, which does not satisfy the Jacobi 
relations, so the imaginary part (\ref{imaginary_part}) is necessary. 
One may reverse its sign,
but given the action there seems to be little, if any, further freedom in the 
bracket. Adding boundary terms to the action, which does not affect the
Peierls bracket, might alter $\{\cdot,\cdot\}_\bullet$.

The data $s_A$, $A = L,R$ seem unphysical as they do not affect the geometry of 
the solution, yet they participate in the Poisson bracket. In fact $s_A$ may be 
replaced almost entirely by $\mu_A$, which commutes with everything. $\mu$ and 
$\mu_A$ together determine $s_A: \theta \mapsto y_A$ up to the three parameter 
group of conformal maps of the unit disk to itself. Moreover $\mu_A$ can always be set 
to zero by a suitable choice of $y_A$ chart. The remaing three degrees of freedom 
in $s_A$ are determined by the boundary values of $s_A$ on $\di S_0$, which 
commute with all data on the interior of $\cN$. Of course, were $S_0$ a 2-sphere 
instead of a disk no boundary values would be available to fix the conformal
automorphisms.






$\rho_0$, $s_L$, $s_R$ and $\mu$ qualify as ``configuration variables'' since 
they form a maximal commuting set among functionals of the data. [We regard $\rho_L(y_L)$ 
and $\rho_R(y_R)$, which commute with everything, as fixed].
A quantization may thus be attempted in terms of wave functionals depending on 
$\rho_0$, $s_L$, $s_R$, and $\mu$, but annihilated by $\dg/\dg \bar{\mu}$.





\begin{acknowledgments}

I thank R. Gambini and C. Rovelli for key discussions, and L. Smolin, R. Epp, 
P. Aichelburg, A. Perez, J. Zapata, L. Freidel, J. Stachel, A. Rendall, 
H. Friedrich, C. Kozameh, G. Aniano, and I. Bengtsson for questions, comments, 
and encouragement. I also thank the CPT in Luminy, AEI in Potsdam, and PI in 
Waterloo, for their hospitality. This work has been partly supported by 
Proyecto PDT 63/076.

\end{acknowledgments}

\bibliography{nullrefsPRL}

\end{document}